\documentclass[11pt,a4paper]{article}
\usepackage{jheppub}
\usepackage{axodraw}
\usepackage{bbm} 
\allowdisplaybreaks 
\usepackage{amsmath} 

\usepackage{caption}
\usepackage{subcaption}

\newcommand{\la}[1]{\label{#1}}
\newcommand{\fig}{Fig.~}
\newcommand{\eq}{Eq.~}
\newcommand{\se}{Sec.~}
\newcommand{\app}{App.~}
\newcommand{\eqs}{Eqs.~}
\newcommand{\nr}[1]{(\ref{#1})}
\newcommand{\nn}{\nonumber \\}
\renewcommand{\(}{\left(}
\renewcommand{\)}{\right)}

\newcommand{\lk}{\left[}
\newcommand{\rk}{\right]}
\newcommand{\ld}{\left.}
\newcommand{\rd}{\right.}
\newcommand{\e}{\epsilon}
\newcommand{\sumint}[1]{\hbox{$\sum$}\!\!\!\!\!\!\!\int_{#1}}

\newcommand{\gammaE}{{\gamma_{\small\rm E}}}

\newcommand{\Nf}{N_{\mathrm{f}}}
\newcommand{\Nc}{N}

\newcommand{\gR}{g_{\mathrm{R}}}
\newcommand{\gB}{g_{\mathrm{B}}}

\newcommand{\gE}{g_{\mathrm{E}}}
\newcommand{\mE}{m_{\mathrm{E}}}

\newcommand{\lone}{\lambda_{\mathrm{E}}^{(1)}}
\newcommand{\ltwo}{\lambda_{\mathrm{E}}^{(2)}}
\newcommand{\mel}{m_{\mathrm{el}}}
\newcommand{\gren}{\frac{\gR^2}{(4\pi)^2}}
\newcommand{\gEr}{g_{\mathrm{ER}}}
\newcommand{\gEb}{g_{\mathrm{E}}}
\newcommand{\mEr}{m_{\mathrm{ER}}}
\newcommand{\mEb}{m_{\mathrm{E}}}
\newcommand{\loneb}{\lambda^{(1)}_{\mathrm{E}}}
\newcommand{\loner}{\lambda^{(1)}_{\mathrm{ER}}}
\newcommand{\ltwob}{\lambda^{(2)}_{\mathrm E}}
\newcommand{\ltwor}{\lambda^{(2)}_{\mathrm{ER}}}
\newcommand{\lbothb}{\lambda^{(1/2)}_{\mathrm E}}
\newcommand{\lbothr}{\lambda^{(1/2)}_{\mathrm{ER}}}
\newcommand{\lboth}{\lambda^{(1/2)}_{\mathrm E}}

\newcommand{\order}[1]{{\mathcal O}(#1)}

\newcommand{\tinymsbar}{{\overline{\mbox{\tiny\rm{MS}}}}}
\newcommand{\msbar}{{\overline{\mbox{\rm{MS}}}}}
\newcommand{\Lambdamsbar}{{\Lambda_\tinymsbar}}
\newcommand{\aEfour}{\alpha_{{\mathrm E}4}}
\newcommand{\aEsix}{\alpha_{{\mathrm E}6}}
\newcommand{\aEeight}{\alpha_{{\mathrm E}8}}
\renewcommand{\vec}[1]{{\mathbf{#1}}}

\newcommand{\qz}{Q_0}

\newcommand{\intV}{J_{11}}
\newcommand{\intVb}{J_{12}} 
\newcommand{\intMcb}{J_{13}}
\newcommand{\Ja}{J_{210011}^{000}} 
\newcommand{\Jb}{J_{220011}^{002}} 
\newcommand{\Jc}{J_{310011}^{020}} 
\newcommand{\Jd}{J_{310011}^{200}}
\newcommand{\Je}{J_{410011}^{130}} 
\newcommand{\Jf}{J_{510011}^{600}} 
\newcommand{\Jg}{J_{530011}^{640}} 
\newcommand{\Jh}{J_{620011}^{730}} 
\newcommand{\Jaa}{J_{111110}^{000}}
\newcommand{\Jab}{J_{211110}^{020}}
\newcommand{\Jac}{J_{31111-2}^{000}}
\newcommand{\Icaa}{I_3 I_1 I_1}
\newcommand{\Ibba}{I_2 I_2 I_1}
\newcommand{\Ica}{I_3 I_1}
\newcommand{\Ibb}{I_2 I_2}
\newcommand{\Iba}{I_2 I_1}
\newcommand{\Iaa}{I_1 I_1}

\newcommand{\picVa}{\begin{picture}(80,40)%
\BCirc(20,20){20}\Line(20,0)(6,34)\Line(20,0)(34,34)%
\end{picture}}
\newcommand{\picVb}{\begin{picture}(80,40)%
\BCirc(20,20){20}\Line(20,0)(6,34)\Line(20,0)(34,34)%
\put(20,40){\circle*{5}}\Line(38,18)(42,22)\Line(38,22)(42,18)%
\end{picture}}
\newcommand{\picMcb}{\begin{picture}(80,40)%
\BCirc(20,20){20}\Line(20,0)(20,20)\Line(20,20)(6,34)\Line(20,20)(34,34)%
\put(12,38.4){\circle*{5}}\put(28,38.4){\circle*{5}}%
\Line(18,10)(22,14)\Line(18,7)(22,11)%
\end{picture}}

\newcommand{\mercedes}{\begin{picture}(30,30)(0,10)%
\CArc(15,15)(15,0,360) \Line(15,0)(15,15)%
\Line(15,15)(25.5,25.5) \Line(15,15)(4.5,25.5)%
\Text(15,35)[]{\text{\scriptsize{$a$}}} \Text(0,3)[]{\text{\scriptsize{$b$}}}%
\Text(30,3)[]{\text{\scriptsize{$c$}}} \Text(7.5,15.5)[]{\text{\scriptsize{$d$}}}%
\Text(22.5,15.5)[]{\text{\scriptsize{$e$}}} \Text(18,6.5)[]{\text{\scriptsize{$f$}}}%
\end{picture}}

\title{Debye screening mass of hot Yang-Mills theory to three-loop order}

\preprint{HIP-2015-24/TH}

\author[a]{Ioan Ghi\c{s}oiu,}
\author[b]{Jan M\"oller}
\author[c]{and York Schr\"oder}

\affiliation[a]{Department of Physics and Helsinki Institute of Physics, P.O.Box 64, 
FI-00014 University of Helsinki, Finland}
\affiliation[b]{Arburg Maschinenbau, 72290 Lo{\ss}burg, Germany}
\affiliation[c]{Grupo de F\'isica de Altas Energ\'ias, Universidad del
B\'io-B\'io, Casilla 447, Chill\'an, Chile}

\emailAdd{ioan.ghisoiu@helsinki.fi}
\emailAdd{jmoeller@physik.uni-bielefeld.de}
\emailAdd{yschroeder@ubiobio.cl}

\abstract{Building upon our earlier work, we compute a Debye 
mass of finite-temperature Yang-Mills theory to three-loop order.
As an application, we determine a $g^7$ contribution
to the thermodynamic pressure of hot QCD.}

\begin{document}
\maketitle
\flushbottom

%
\section{Introduction}
\la{se:intro}

In electromagnetic plasmas, the Debye screening 
mass $\mel$ --- or the inverse 
screening length of electric fields within the plasma --- is a most
fundamental quantity.  While it is sometimes defined as the 
small-momentum limit of the static Coulomb propagator 
$1/[\vec k^2+\Pi_{00}(0,\vec k)]$
as 
$m_{\rm D}^2=\Pi_{00}(\omega=0,\vec k\!\rightarrow\!0)$, 
where $\Pi_{00}(\omega,\vec k)$ is the longitudinal part of the
photon self-energy, an alternative definition
is in terms of the pole of the same static 
propagator,
\begin{align}
\la{eq:gap}
0 &= \vec k^2 + \Pi_{00}(0,\vec k) \big|_{\vec k^2=-\mel^2} \;.
\end{align}
Equivalently, in a quark-gluon plasma, the Debye screening mass 
parameterizes the dynamically generated screening of chromo-electric
fields, due to the strong interactions of hot quantum chromodynamics (QCD).
Now, since the longitudinal gluon self-energy is not a gauge 
invariant quantity, and since its static low-momentum limit
exhibits severe infrared (IR) divergences \cite{Landsman:1986uw,Linde:1980ts}, 
it is not at all obvious
that the above definitions for a physical quantity make sense.
Indeed, after a number of investigations of this matter
\cite{Kajantie:1981hu,Kajantie:1982xx,Furusawa:1983gb,Toimela:1982ht,Arnold:1995bh},
it has turned out that the definition of \eq\nr{eq:gap} is
the physically sensible one\footnote{Note that the proper definition of the pole
location is a bit more subtle, see e.g.\ \se5 
of \cite{Bieletzki:2012rd}, but this makes no difference here.}, 
leading to a gauge invariant and
infrared finite Debye mass also at higher orders in perturbation
theory, as has been demonstrated in 
Refs.~\cite{Kobes:1990xf,Kobes:1990dc,Rebhan:1993az,Braaten:1994pk}.

The analytic treatment of such hot QCD systems can be transparently
organized after identifying the different dynamically generated 
energy scales.
Applying the concept of effective theories (EFTs) 
to this multi-scale system, it has been understood how to reduce
the root cause of the IR problem to a well-defined non-perturbative
lattice measurement which, after mapping to the continuum,
constitutes a systematic approach that evades the IR problem 
and renders thermodynamic observables theoretically computable, 
allowing for systematic improvements.
In the case of QCD in thermal equilibrium, one can identify three
relevant energy scales, being effectively described by a set of
three distinct theories: 4-dimensional hot QCD, 
3-dimensional Electrostatic QCD (EQCD) and 
3-dimensional Magnetostatic QCD (MQCD), respectively. 
Together, after proper matching of their
parameters, they allow for consistent weak-coupling expansions 
of static quantities.

In the present paper, by a three-loop determination of the Debye mass
of hot Yang-Mills theory,
we intend to contribute yet another coefficient 
to the weak-coupling EFT setup, which is needed for matching 
QCD and EQCD. 
Our motivation is threefold.
First, our new result immediately determines a (gauge invariant part of the)
$g^7$ contribution to the thermodynamic pressure.
The importance of this higher-order contribution lies in the fact 
that it represents the next-to leading order 
correction to a {\em physical leading order}
(LO), ultimately enabling the first sound statements about convergence 
and (renormalization) scale dependence, going beyond existing 
discussions that are based on truncated (or, in the modern understanding,
incomplete\footnote{Since the effect of large logarithms,
a well-known effect in EFTs every time a new physical scale enters the problem,
is then not considered properly.} LO) versions of the series.

Second, the Debye mass plays a prominent role in various channels
of gauge-invariant gluonic screening masses  \cite{Arnold:1995bh,Hart:2000ha,Laine:2009dh}.
Within a weak-coupling expansion, a number of these are at leading order
proportional to the Debye mass followed by a formally sub-leading, 
but numerically large, logarithmic correction.
There are examples, however, where the logarithmic term is known to be small, 
such that the Debye mass dominates the functional behavior in the phenomenologically
relevant temperature range, such as has been observed for the lowest-lying 
color-magnetic screening mass \cite{Laine:2009dh}.

Third, we regard the determination of the 3-loop Debye screening
mass $\mE^2$, as an EFT matching parameter, as a proof-of-principle
that also the corresponding evaluation of the 3-loop effective gauge 
coupling constant $\gE^2$ is within reach.
The latter does bring a direct phenomenological
application, allowing for a precise comparison of the so-called
spatial string tension $\sigma_s$ (as evaluated in the EFT setting)
with lattice determinations, where a previous 2-loop analysis
had shown great promise, while leaving room for further corrections.
Furthermore, such a comparison can be regarded as an important consistency check,
validating the EFT approach as a whole.

The structure of the paper is the following. 
In \se\ref{se:setup} we provide a brief overview of the theoretical setup
and of the matching relations between full QCD and EQCD and sketch the status
of the Debye mass computation reached in Ref.~\cite{Moeller:2012da}. 
In \se\ref{se:eval} we complete the calculation, restricting ourselves
to the case $N_f=0$ (pure gauge theory), express the bare
mass parameter in terms of a few master sum-integrals, renormalize
and analyze our result numerically.
We then use the renormalized result for
extracting a $g^7$ contribution
to the QCD pressure in \se\ref{se:g7}.
We conclude in \se\ref{se:conclu}, while some technicalities 
are relegated to the appendices.

%
\section{Setup: effective theory and matching}
\la{se:setup}

At finite temperatures, gluons exhibit three characteristic  
momentum scales 
($2\pi T$, $g T$ and $g^2 T$) of which the  {\em ultra-soft}
color-magnetic mode ($g^2 T$) leads to a breakdown of the 
ordinary perturbative expansion \cite{Linde:1980ts}.
A straightforward approach that sidesteps this problem is scale separation. 
This is achieved by constructing dimensionally reduced effective
theories \cite{Ginsparg:1980ef,Appelquist:1981vg,Kajantie:1995dw} 
whose parameters are matched to full QCD. 
Following this procedure, two effective theories in $d=3$ dimensions emerge:
Electrostatic QCD (EQCD) is an $SU(N)$ gauge theory coupled to a massive
scalar field in the adjoint representation. 
EQCD contains two scales (the {\em soft} and the {\em ultra-soft} one)
while the {\em hard} scale enters only through the perturbative matching 
of the theory's parameters.
Magnetostatic QCD (MQCD) is a pure $SU(N)$ gauge theory containing only the 
non-perturbative ultra-soft scale,
whereas the hard and the soft scales enter again only through the matched 
parameters of the theory.
MQCD can only be studied with non-perturbative 
methods such as lattice QCD \cite{Hietanen:2004ew,Hietanen:2006rc}.

In the following, we are concerned only with matching hot QCD to EQCD.
The bare EQCD Lagrangian reads
\begin{align}
\la{eq:eqcd_lagr}
\mathcal{L}_{\text{EQCD}}^{3d} &= 
-\frac1{2\gE^2}\mathrm{Tr}[D_i,D_j]^2 
+\mathrm{Tr}[D_i, A_0]^2 
+\mE^2 \mathrm{Tr}A_0^2 
+\lone (\mathrm{Tr}A_0^2)^2
+\ltwo \mathrm{Tr}A_0^4
+\delta\mathcal{L}_{\mathrm E}\;,\nonumber\\
D_i &= {\mathbbm 1}\,\partial_i -i \gE A_i \;,\quad 
A=A^a T^a\;,\quad 
[T^a,T^b]=if^{abc}T^c\;,\quad 
\mathrm{Tr}\,T^aT^b=\delta^{ab}/2\;.
\end{align}
The operators quartic in the fields $A_0$ are linearly independent 
for $\Nc>3$ only, while for $\Nc=2,3$ we have 
$\mathrm{Tr}A_0^4=\frac12(\mathrm{Tr}A_0^2)^2$.
The term $\delta{\cal L}_{\mathrm E}$ collects the infinite tower of higher-order operators
that are generated by integrating out the hard scale, the lowest of which have
been classified in Ref.~\cite{Chapman:1994vk}. We shall be needing a single
one of them, cf.\ \eq\nr{eq:deltaL} below.

The detailed framework of performing the matching computation has been 
presented in \cite{Moeller:2012da,Moller:2012zz}. 
Here, we merely provide a concise version of it and 
generalize the matching condition in order to account for
higher-order operators.
The general prescription is to require that various 
static quantities computed in both theories
match to a certain order in a strict perturbative 
expansion with respect to the gauge coupling $g$. 
By using the background field gauge, we make sure that on the QCD side 
only the coupling constant $g$ requires renormalization \cite{Abbott:1980hw,Abbott:1981ke}.

%
\subsection{Screening mass in QCD}
\la{se:QCDmatch}

Following \eq\nr{eq:gap},
we define the screening mass $\mel$ as the pole of the 
static ($K_0 = 0$) momentum-space propagator of $A_0$ with the 
on-shell condition $\vec k^2 = -\mel^2$.
On the full QCD side, writing 
$\Pi_{00}^{ab}(0,\vec k)\equiv\delta^{ab}\Pi_{\mathrm E}(0,\vec k)$
we therefore have
\begin{align}
\la{eq:QCDpole}
0=\ld \vec k^2+ \Pi_{\mathrm E}(0,\vec k)\right|_{\vec k^2 = -\mel^2} \;.
\end{align}
The self-energy is written as an expansion 
in both the gauge coupling $g$ and in the external 
momentum $\vec k$ in order to permit a strict perturbative expansion. 
The $\vec k$\/-expansion is justified due to the soft 
scale $|\vec k| \propto gT$ at which the pole in the propagator
 arises \cite{Braaten:1995jr}:
\begin{align}
\la{eq:Piexpansion}
\Pi_{\mathrm E}(0,\vec k) &= 
\sum_{n=1}^{\infty}(g^2)^n\Big[\Pi_{{\mathrm E}n}+\vec k^2\,\Pi_{{\mathrm E}n}'
+(\vec k^2)^2\,\Pi_{{\mathrm E}n}''+\dots\Big]\;.
\end{align}
Solving iteratively, \eq\nr{eq:QCDpole} leads to the following expression 
for the screening mass:
\begin{align}
\la{eq:mel}
\mel^2 &= 
 g^2\, \Pi_{\mathrm E1}
+g^4 \lk \Pi_{\mathrm E2}-\Pi_{\mathrm E1}'\Pi_{\mathrm E1} \rk
+\nonumber\\&+
 g^6\lk
 \Pi_{\mathrm E3} 
-\Pi_{\mathrm E1}' \Pi_{\mathrm E2} 
-\Pi_{\mathrm E2}' \Pi_{\mathrm E1} 
+\Pi_{\mathrm E1}''(\Pi_{\mathrm E1})^2
+\Pi_{\mathrm E1} (\Pi_{\mathrm E1}')^2 
\rk +\order{g^8}\;.
\end{align}
Full $d$\/-dimensional representations of the various 
coefficients $\Pi_{\mathrm En}$ can be found 
in \app{C} of \cite{Moeller:2012da}. 
For the reader's convenience, we have collected the 
corresponding one- and two-loop self-energies in \app\ref{se:lower} below.

The evaluation of the QCD self-energy tensor $\Pi_{\mu \nu}(K)$ to three-loop
order generates approximately 500 Feynman diagrams, necessitating an automatized
procedure to handle this task 
(cf.\ \cite{Moeller:2012da,Moller:2012zz} and references therein). Generation of the Feynman
diagrams, the color algebra computation of $SU(N)$, the Lorentz contraction
and the Taylor expansion into external momentum have been performed using specialized
software (we have employed QGRAF \cite{Nogueira:1991ex,Nogueira:2006pq} 
and FORM \cite{Vermaseren:2000nd,Kuipers:2012rf}). 
The resulting $\approx 10^7$ sum-integrals have then been reduced 
via systematic use of integration-by-parts (IBP) relations \cite{Laporta:2001dd} 
in the thermal context \cite{Nishimura:2012ee}
to a small number of master sum-integrals
$J$ and $I$, as defined in \app\ref{se:masters}.

For $N_{\text f}=0$, the bare 3-loop result 
of Refs.~\cite{Moeller:2012da,Schroder:Reduction} reads
\begin{align}
\la{eq:PiOld}
\Pi_{\mathrm E3} &= \Nc^3\Big( 
\sum_{i=1}^8 \alpha_i(d)\,J_i
+\alpha_9(d,\xi)\, \Icaa
+\alpha_{10}(d,\xi)\, \Ibba \Big) \;,\\
\la{eq:Jmasters}
\big\{J_1,\dots,J_8\big\} &\equiv
\big\{\Ja,\Jb,\Jc,\Jd,\Je,\Jf,\Jg,\Jh\big\} \;,
\end{align}
with pre-factors $\alpha_{1...10}$ given in \app\ref{se:coeffs}.
Inspecting this result, there are two trivial products of
one-loop tadpole sum-integrals $\sim I\cdot I\cdot I$ that are
already known analytically (cf.\ \eq\nr{eq:Idef})
and eight non-trivial three-loop cases $J_{ab00cd}^{\alpha\beta\gamma}$ 
of basketball-type, of which so far only a few terms of 
the one multiplying $\alpha_1$ (originally given in \cite{Gynther:2007bw}
and re-evaluated as a specific case of the class $J_{N10011}^{000}$
in \cite{Moller:2010xw}) as well as the one
multiplying $\alpha_4$ (treated as a special
case of $J_{N10011}^{M00}$ in \se5 of \cite{Moeller:2012da}) are known.

As discussed in Ref.~\cite{Moeller:2012da}, the difficulty in 
calculating the 3-loop sum-integrals $J_i$ that appear in \eq\nr{eq:PiOld} 
lies in the fact they would have to be expanded beyond the constant term in 
$\e$ (in our notation, $d=3-2\e$) 
due to their singular pre-factors,
cf.\ \app\ref{se:coeffs}.
Conventional techniques of evaluating basketball-type 
sum-integrals \cite{Arnold:1994ps,Arnold:1994eb}, relying on setting $d=3$ for determining 
their constant parts in coordinate space representations,
make this task difficult (if not impossible).
We will show in \se\ref{se:eval} how to proceed.

%
\subsection{Screening mass in EQCD}

In EQCD, writing the self-energy of the adjoint scalar $A_0$ as 
$\Pi_{\text{EQCD}}^{ab}=\delta^{ab}\,\Pi_{\text{EQCD}}$, 
the pole mass of the $A_0$ propagator is 
\begin{align}
\la{eq:EQCDpole}
0=\ld \vec k^2 + \mE^2 + \Pi_{\text{EQCD}}(\vec k)\right|_{\vec k^2 = -\mel^2} \;.
\end{align}
By performing the same twofold expansion of the $A_0$ self-energy $\Pi_{\text{EQCD}}$ 
as in \eq\nr{eq:Piexpansion}, it vanishes
to all orders in perturbation theory due to the absence of any scale
in the resulting three-dimensional vacuum 
integrals\footnote{Regarding the mass term $\mE \propto g T$ as a perturbation,
the propagator can be taken to be massless.}
(cf.\ discussion in \se 2.1 of \cite{Moeller:2012da}).
However, in \cite{Moeller:2012da} higher-order operators to \eq\nr{eq:eqcd_lagr},
such as those computed in \cite{Chapman:1994vk} and that contribute to
$\order{g^6}$ in $\mE^2$, were not yet considered. 
Indeed, in the context of the double expansion, in which only scale-free 
vacuum integrals emerge, one might expect that
these higher-order contributions all vanish.

It turns out, however, that one higher-order operator does contribute 
to the self-energy, since it generates a tree-level two-point
contribution that is not affected by the momentum expansion.
This dimension six operator can be extracted from Ref.~\cite{Chapman:1994vk}:
\begin{align}
\la{eq:deltaL}
\delta \mathcal{L}_{\mathrm E} &\ni
c_6 \times \gR^2\,\mathrm{Tr} (\partial_i^2 A_0 )^2 \;,\quad
c_6 = \frac{34 \Nc}{15}\,\frac{\zeta(3)}{(4\pi)^4\,T^2}\;,
\end{align}
where $\gR^2$ is the (dimensionless) renormalized 4d QCD gauge coupling.
Since at the pole, the four derivatives will scale as 
$(\vec k^2)^2\sim g^4T^4$,
\eq\nr{eq:EQCDpole} receives a correction 
of $\order{g^6}$.
Adding the contribution of \eq\nr{eq:deltaL} to $\Pi_{\mathrm{EQCD}}$, it reads
\begin{align}
\la{eq:melEQCD}
0=-\mel^2 +\mE^2 +c_6 \,\gR^2(-\mel^2)^2 +\order{g^8} \;.
\end{align}
The matching now follows from replacing $\mel^2$ in \eq\nr{eq:melEQCD} with its 
QCD value \eq\nr{eq:mel}.

%
\section{Completing the calculation}
\la{se:eval}

As already emphasized, a technically challenging part
of the matching is the evaluation of the master sum-integrals.
We also mentioned the impossibility of computing sum-integrals beyond the
constant term in $\e$ with state of the art techniques.
However, in \eq\nr{eq:PiOld} 
the singular pre-factors in $\e$ multiplying the master sum-integrals
require their evaluation including $\order{\e}$.

%
\subsection{Basis transformation}
\la{se:evalbasisTrafo}

A possible way out is to search for a suitable basis transformation that
-- much in the spirit of the $\e$\/-finite basis advocated in
Ref.~\cite{Chetyrkin:2006dh} -- removes the singularities of 
the pre-factors in \eq\nr{eq:PiOld} for the price 
of introducing master sum-integrals that are of a different
topology and might therefore be more difficult to compute. 
Using the master integrals defined in \app\ref{se:masters}
as well as the IBP relations of \app\ref{se:ibp}, we have succeeded in 
rewriting the bosonic part of $\Pi_{\mathrm E3}$ shown in \eq\nr{eq:PiOld}
as the remarkably compact expression
\begin{align}
\la{eq:PiNew}
\frac{\Pi_{\mathrm E3}}{\Nc^3(d-1)^2} &= 
-\frac{d-3}{4}\Big[
(7d-13)\,\intV
+32(d-4)\,\intVb
+2(d-7)\,\intMcb 
\Big]
+\nn&+
\Big[\frac{(d-7)(3d-7)}{(d-1)^2}+\xi+\frac{d-6}{12}\,\xi^2\Big](d-1)^2\Icaa
+\\&+
\Big[\frac{85\!+\!4d\!-\!15d^2\!+\!2d^3}{d-5}+(16\!-\!13d\!+\!2d^2)\,\xi
+\frac{16\!-\!13d\!+\!3d^2}{4}\,\xi^2\Big]\frac{d-3}{2(d\!-\!2)}\,\Ibba
\;,\nn
\la{eq:newBasis}
\big\{\intV,\intVb,\intMcb\big\} &\equiv
\big\{\Jaa,\Jab,\Jac\big\}\;.
\end{align}
Using the lower-order self-energies listed in \app\ref{se:lower},
\eq\nr{eq:mel} then immediately gives 
\begin{align}
\la{eq:melbare}
m_{\mathrm{el}}^2 &= \Nc g^2 (d-1)^2 I_1 \,\bigg\{ 1 
+ \Nc g^2 \frac{46 - 11 d + d^2}{6}\, I_2 
+\nn&+ 
\Nc^2 g^4 \bigg( -\frac{d-3}4\,\Big[ (7 d -13) \intV /I_1
+32 (d -4) \intVb /I_1
+2(d -7) \intMcb /I_1 \Big]
+\nn&+
\frac1{6d(d-7)}\,\Big[ \frac{p_1(d)}5\,\Ica 
+\frac{p_2(d)}{6(d-5)(d-2)}\,\Ibb \Big] \bigg)
+\order{g^6}\bigg\}\;,
\end{align}
with polynomials
$p_1(d)=(720 - 12472 d + 9779 d^2 - 2686 d^3 + 364 d^4 - 26 d^5 + d^6)$
and 
$p_2(d)=(3024 - 42028 d + 81720 d^2 - 56428 d^3 + 19783 d^4 - 3898 d^5 + 
 448 d^6 - 30 d^7 + d^8)$.
Note that all dependence on the gauge parameter $\xi$ has duly canceled
in this $d$\/-dimensional result.
By plugging the leading term of \eq\nr{eq:melbare}
into \eq\nr{eq:melEQCD}, we finally obtain
\begin{align}
\mE^2=\mel^2 -c_6\,\gR^2\,g^4 \Nc^2 (d-1)^4 \Iaa +\order{g^8}\;.
\end{align}

\begin{figure}
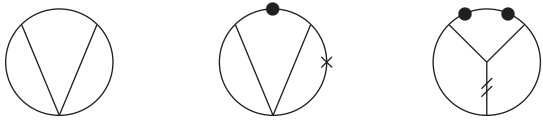

\begin{center}
\picVa
\picVb
\picMcb
\end{center}
\caption{The three non-trivial three-loop sum-integrals $\intV$, $\intVb$ and $\intMcb$
of \eq\nr{eq:newBasis} that are needed for the Debye mass.
A dot on a line stands for an extra power of the corresponding
propagator, a cross denotes an extra factor of $P_0^2$ in the numerator,
and a slash means the line appears in the numerator.}
\la{fig:sumints}
\end{figure}

The remaining task is to evaluate the sum-integrals $J_{11,12,13}$ that enter
our expression for $\mel^2$ and that are depicted in \fig\ref{fig:sumints}.
To this end, the 1-loop substructure can be exploited, a method pioneered 
by Arnold and Zhai \cite{Arnold:1994ps,Arnold:1994eb}.
Their technique of solving basketball-type and spectacle-type 
sum-integrals relies on a careful subtraction of sub-divergences which is 
specific for every sum-integral in part. Nevertheless, it was possible 
to develop a semi-automatized procedure for an analytic calculation 
of the divergent parts of a large class of spectacle-type sum-integrals. 
This was necessary for evaluating the 
Mercedes type sum-integral with two inverse propagators, $\intMcb$. 
Its computation required the use of the dimensional method of 
Tarasov \cite{Tarasov:1996br}, in which the tensor structure of a
sum-integral is translated into a sum of higher dimensional scalar 
integrals.

The last remaining pieces of the matching computation,
the 3-loop vacuum sum-integrals $\intVb$ and $\intMcb$,
have been determined only recently in Refs.~\cite{Ghisoiu:2012kn,Ghisoiu:2012yk}
and are listed in \app\ref{se:masters}.

%
\subsection{Renormalization}
\la{se:res}

We now turn to renormalized quantities.
In the 4-dimensional theory, we need to renormalize the gauge coupling.
The bare coupling $\gB$ (denoted as $g$ in the previous sections) is related to the
renormalized coupling $\gR$ via 
$\gB^2 = \gR^2 \mu^{2\e} Z_g$ (note that $\gR^2$ is dimensionless), $\bar \mu$ 
being the $\msbar$ scheme scale defined as 
$\bar\mu^2 = 4\pi e^{-\gamma_E}\mu^2 $ and
\begin{align}
Z_g = 1 
+ \frac{\gR^2}{(4\pi)^2}\,\frac{\beta_0}{\e} 
+ \frac{\gR^4}{(4\pi)^4}
\( \frac{\beta_1}{2\e} + \frac{\beta_0^2}{\e^2} \) 
+ \frac{\gR^6}{(4\pi)^6}
\( \frac{\beta_2}{3\e} + \frac{7 \beta_0 \beta_1}{6 \e^2} 
+ \frac{\beta_0^3}{\e^3}\)
+\order{g^8}\;,
\end{align}
where, for $\Nf=0$, $\beta_0=-11\Nc/3$, $\beta_1=-34\Nc^2/3$ and $\beta_2=-2857\Nc^3/54$.

In the 3-dimensional theory, only the mass parameter requires renormalization.
This is due to the fact that the EQCD Lagrangian in \eq\nr{eq:eqcd_lagr} 
is super-renormalizable in $d=3-2\e$ dimensions.
The few divergent diagrams in this theory arise solely at two-loop order
and account only for the mass renormalization.
Hence, the renormalization relations are simply
\begin{align}
\gEb^2 = \gEr^2 \mu_3^{2\e}\;,\quad
\lbothb = \lbothr \mu_3^{2\e}\;,\quad
\mEb^2 = \mEr^2 + \delta \mE^2\;,
\end{align}
the renormalized EQCD couplings $\gEr^2$ and $\lbothr$ being of dimension one,
while the EQCD mass counterterm reads \cite{Rajantie:1997pr,Laine:1997dy,Farakos:1994kx}
\begin{align}
\la{eq:dmE}
 \delta \mE^2 &= -\frac{1}{(4\pi)^2} \frac{1}{4\e} \lk 2\Nc \gEr^2 \( (\Nc^2+1) \loner + (2 \Nc^2-3) \frac{\ltwor}{\Nc}\) \rd \nn
 & \ld -2(\Nc^2+1) [\loner]^2 -4(2\Nc^2-3) \loner \frac{\ltwor}{\Nc} -(\Nc^4-6\Nc^2+18) \frac{[\ltwor]^2}{\Nc^2} \rk\;.
\end{align}
This leads to an exact RG equation for the mass parameter: 
$\mu_3^2\partial_{\mu_3^2}\,\mEr^2 = 2\e\,\delta \mE^2$.

In order to express $\delta \mE^2$ in terms of the 4d QCD coupling $\gR^2$,
the EQCD parameters $\gE^2$ and $\lboth$ need to be matched to full QCD at leading order (1-loop).
The corresponding relations are given in the literature as $\gEr^2 = \gR^2 T$,
$\loner = \gR^4 T /(4 \pi^2)$ and $\ltwor = \gR^4 T \Nc/(12 \pi^2)$ \cite{Appelquist:1981vg,Landsman:1989be},
however without regularization parameter $\e$, which we here need due to the presence
of the $1/\e$ divergence in \eq\nr{eq:dmE}.
We hence need $d$\/-dimensional matching relations,
which can be extracted from \cite{Kajantie:1995dw}, in our notation, as
\begin{align}
\gEb^2 = \gB^2 T +\order{\gB^4}\;,\;
\loneb = \frac{(3\!-\!d)(d\!-\!1)^2}{2} I_2 \gB^4 T +\order{\gB^6}\;,\;
\ltwob = \frac{\Nc}3\,\loneb +\order{\gB^6}\;.
\end{align}

Using these expressions, the mass counterterm simplifies to
\begin{align}
\delta \mE^2 = \frac{5\Nc^3(d-3)(d-1)^2 \mu^{2\e}I_2}{12\e}  \(\frac{\mu}{\mu_3}\)^{4\e} \frac{\gR^6 T^2}{(4\pi)^2}
+\order{\gR^8 T^2}\;,
\end{align}
such that we finally get the renormalized Debye mass up to 3-loop order as
\begin{align}
\la{eq:final}
\frac{\mEr^2}{(4\pi T)^2} &= 
 \frac{\Nc}{3} \gren  
+ \lk \frac{\Nc}{3} \gren \rk^2 \( 22 L +5\)+ \nn
&+ \lk \frac{\Nc}{3} \gren \rk^3 \(  484 L^2 +424 L -180 L_3 + \frac{731}{2} -\frac{56\zeta(3)}{5} -3 \times \frac{34\zeta(3)}{15}  \)  +  \order{\gR^8} \;, \nn
 L &= \ln \frac{\bar \mu e^{\gamma_{\mathrm E}}}{4 \pi T}\;,\;\; L_3= \ln \frac{\mu_3^2 e^{Z_1}}{4 \pi T^2}\;,\;\; Z_1 = \frac{\zeta{'}(-1)}{\zeta(-1)}\;,
\end{align}
where the 1-loop and 2-loop contributions have been calculated in 
Ref.~\cite{Laine:2005ai} and where we have expressed the contribution coming 
from \eq\nr{eq:deltaL} separately.

The $\gR^6$ coefficient contains two independent mass scales, $\mu$ and $\mu_3$.
The first arises through the 4d dimensional regularization scheme and the subsequent
renormalization of the 4d coupling, $\gR$, whereas the second scale $\mu_3$
enters through the regularization of the divergent integrals of the 3d $SU(N)$ $+$
adjoint Higgs theory and ultimately through the mass renormalization.

%
\subsection{Numerical evaluation}
\la{se:numev}

For the numerical evaluation of $\mEr$, the running of the 4d coupling
with respect to the energy scale is obtained by solving the RGE iteratively 
to three-loop order \cite{Agashe:2014kda,Chetyrkin:1997sg}
\begin{align}
\la{eq:RGErunning}
\gren = - \frac{1}{\beta_0 t} - \frac{\beta_1 \ln t}{\beta_0^3 t^2} - \frac{1}{\beta_0^3 t^3} \( \frac{\beta_1^2}{\beta_0^2} (\ln^2 t - \ln t-1) + \frac{\beta_2}{\beta_0}\)\;,
\end{align}
with $t = \ln [\bar \mu^2/ \Lambda_{\overline{\text{MS}}}^2]$ and $\Lambda_{\overline{\text{MS}}}$ the QCD scale defined in the $\msbar$ scheme \cite{Bardeen:1978yd,Chetyrkin:1997sg}.

In order to display numerical results, we need to choose values for the two arbitrary mass scales,
$\mu$ and $\mu_3$. For the former one, we adopt the procedure of 
of minimal sensitivity \cite{Laine:2005ai}. 
The scale is computed to be $\bar \mu_{\text{opt}}/T =4 \pi e^{-\gammaE-\frac{1}{22}} \approx 2\pi$.
We then extend this idea to choosing $\mu_{3,\text{opt}}$.
As $\mEr^2(\bar \mu,\mu_3)$ is not a monotonic function with respect to $\bar \mu$,
we impose that the absolute variation of $\mEr^2(\bar \mu,\mu_3)$ in the interval
$(\bar \mu_{\text{opt}}/2,2 \bar \mu_{\text{opt}})$ is minimal for a specific scale
$\mu_3 \equiv \mu_{3,\text{{opt}}}$, or
\begin{align}
 \lk\frac{\partial}{\partial \mu_3} \int_{\frac{\bar \mu_{\text{opt}}}{2}}^{2 \bar \mu_{\text{opt}}} \left| \frac{\partial \mEr^2(\bar \mu, \mu_3)}{\partial \bar \mu} \right| \mathrm d \bar \mu\rk_{\mu_3 =\mu_{3,\text{opt}}} =0\,.
\end{align}
Taking the absolute variation ensures that an oscillatory behavior of $\mEr^2$ in
the considered interval is ruled out to be regarded as a scale for minimal sensitivity.
Solving the equation numerically, we obtain $\mu_{3,\text{opt}}/T \approx 2.85$.

\begin{figure}
\centering
\begin{subfigure}[b]{0.45\textwidth}
\includegraphics[scale=0.57]{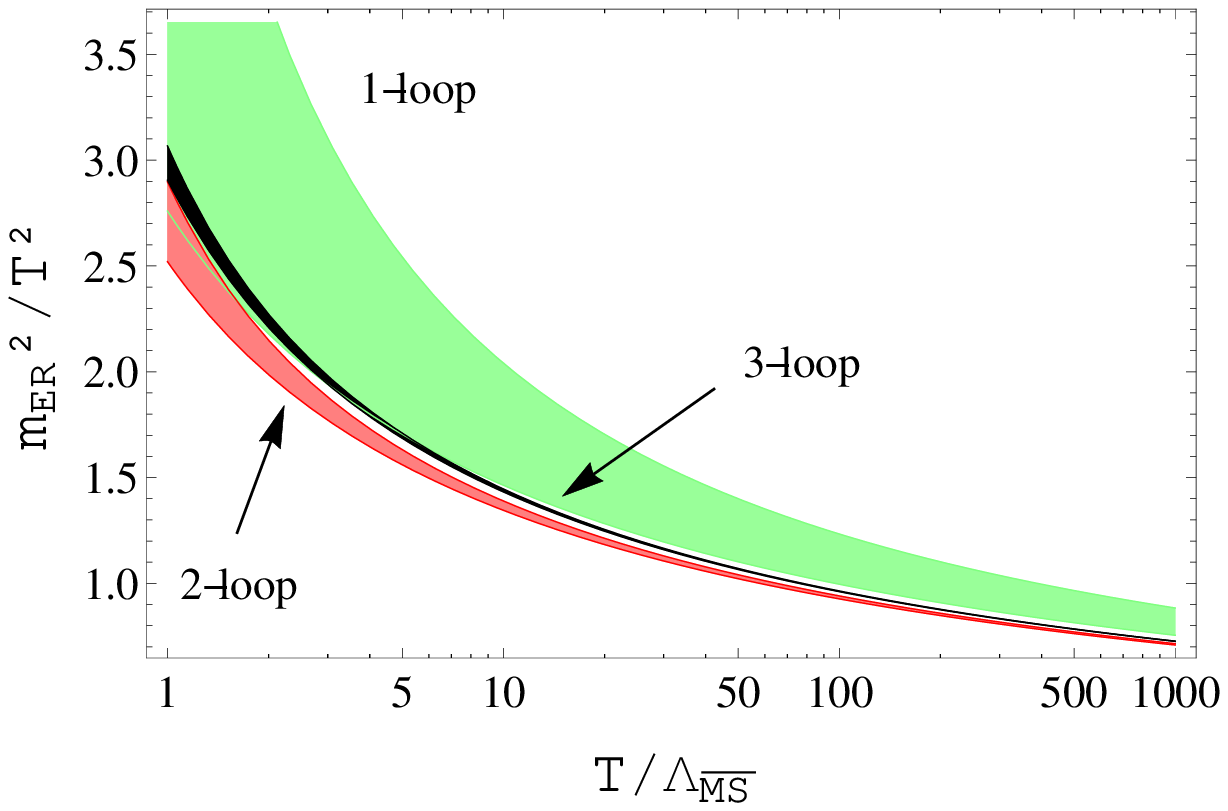} 
\end{subfigure}
\hspace{5mm}
\begin{subfigure}[b]{0.45\textwidth}
\includegraphics[scale=0.56]{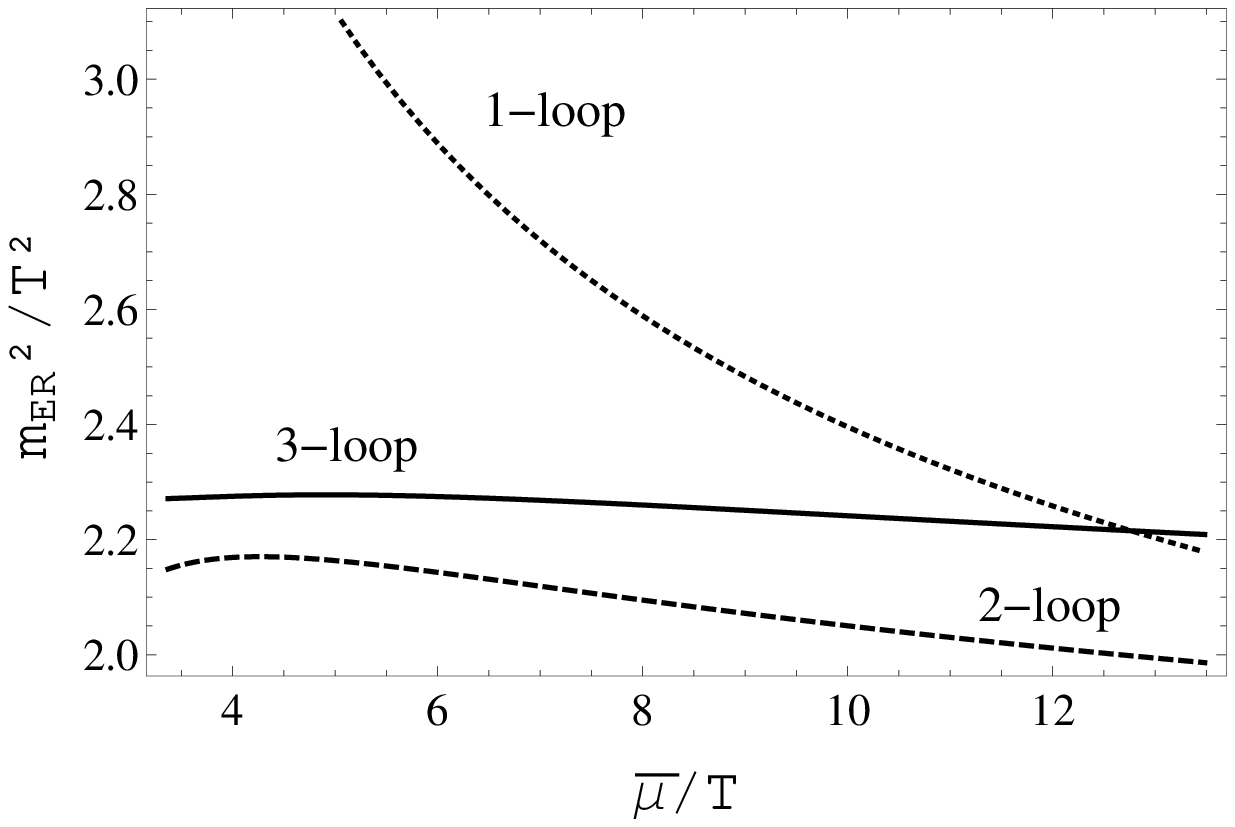}
\end{subfigure}
\caption{Left: dimensionless Debye mass $\mEr^2/T^2$ as a function 
of the temperature $T$, in units of $\Lambdamsbar$ and with the variation of $\bar \mu = (0.5 ... 2.0) \times \bar \mu_{\text{opt}}$. Right: variation of the mass with respect to $\bar \mu$ at the scale $T/\Lambdamsbar = 2$ and with $\mu_3 = \mu_{3,\text{opt}}$.}
\la{fig:mE}
\end{figure}

In \fig\ref{fig:mE}, we
analyze the running of the Debye mass with respect to the 
temperature ($T$).
We have used the solution of the 
renormalization group equation 
of the 4d coupling $g$, in which the QCD $\beta$-function was 
truncated after $\order{g^{8}}$. The parameter $\Lambdamsbar$ 
corresponds to the QCD scale defined in Ref.~\cite{Laine:2005ai},
and here simply sets the scale through the $t$\/-dependence of the running coupling \eq\nr{eq:RGErunning}.
The arbitrary scale $\bar \mu$ was chosen at the point where 
the effective coupling $\gEr$ has a minimal sensitivity 
to it: $\bar \mu_{\text{opt}}/T \approx 2\pi$. 
The 3 bands in the plot arise by varying
$\bar \mu = (0.5 ... 2) \times \bar \mu_{\text{opt}}$. Using the prescription described above for choosing a sensible scale for the EQCD scale parameter, $\mu_3/T \approx 2.85$, we obtain a three-loop result with a vanishingly narrow band width.

From the figure, one notices a slight increase of the Debye mass with respect to the 2-loop
result. In addition, the sensitivity with respect to the arbitrary scale $\bar
\mu$ decreases, which indicates that the perturbative expansion up to 3-loop
order shows good convergence properties.

%
\section{A $g^7$ contribution to the QCD pressure}
\la{se:g7}

Having the 3-loop Debye mass at hand, we can use it to extract a gauge-invariant piece of
a higher-order perturbative ($g^7$) correction to the QCD pressure. The order $g^7$ owes 
its importance to the fact that it represents, in the effective theory setup we are working in, 
the leading correction to what has been called the {\em physical leading order}, i.e.\ all terms
up to order $g^6$ that, for the first time, include all potentially large logarithms entering the
QCD pressure \cite{Kajantie:2002wa}.

As already discussed in the introduction, 
in the effective theory framework, the QCD partition function factorizes, such that
the pressure $p_{\mathrm{QCD}}$ splits 
into three parts $p_{\mathrm{E}},\,p_{\mathrm{M}}$ and $p_{\mathrm{G}}$ 
originating from contributions of the hard- $2 \pi T$, soft- $g T$ and 
ultra-soft scale $g^2 T$, respectively \cite{Braaten:1995jr,Kajantie:2000iz}. 
These three parts can be extracted as matching coefficients of QCD, EQCD and MQCD, along the following
chain of equations (recall $d=3-2\e$),
\begin{align}
p_{\mathrm{QCD}}(T) 
&= \lim_{V\rightarrow\infty} \frac1{V_{d+1}} 
\ln\int\!\!\mathcal{D}A_\mu^a\;
\exp\Big[-\int_0^{1/T}\!\!\!\!\!\!\!{\rm d}\tau\int\!\!{\rm d}^{d}x\; \mathcal{L}^{\mathrm{4d}}_{\mathrm{QCD}}\Big] \nn
&= p_{\mathrm{E}}(T) + 
\lim_{V\rightarrow\infty} \frac{T}{V_d} 
\ln\int\!\!\mathcal{D}A_k^a\,\mathcal{D}A_0^a\;
\exp\Big[-\int\!\!{\rm d}^dx\; \mathcal{L}^{\mathrm{3d}}_{\mathrm{EQCD}}\Big] \nn
&= p_{\mathrm{E}}(T) + p_{\mathrm{M}}(T) + 
\lim_{V\rightarrow\infty} \frac{T}{V_d} 
\ln\int\!\!\mathcal{D}A_k^a\;
\exp\Big[-\int\!\!{\rm d}^dx\; \mathcal{L}^{\mathrm{3d}}_{\mathrm{MQCD}}\Big] \nn
&= p_{\mathrm{E}}(T)+p_{\mathrm{M}}(T)+p_{\mathrm{G}}(T)\,,
\la{eq:decomp_pres}
\end{align}
supplemented with the corresponding matching of the couplings, as has been explained 
in the previous chapters on the example of the chromo-electric screening mass.

It turns out that the different pieces in \eq\nr{eq:decomp_pres}, when re-expressed in terms
of the renormalized 4d gauge coupling $\gR$ (and omitting logarithms of the coupling), 
contribute as
\begin{align}
p_{\mathrm{E}}&\sim T^4(1+\gR^2+\gR^4+\gR^6+\dots)\;,\\
p_{\mathrm{M}}&\sim T^4(\gR^3+\gR^4+\gR^5+\gR^6+\gR^7+\dots)\;,\\
p_{\mathrm{G}}&\sim T^4(\gR^6+\gR^7+\dots)\;.
\end{align}
Both parts of the pressure coming from the 3d reduced effective theories describing
soft and ultra-soft effects, EQCD and MQCD, contain contributions of order $\gR^7$.
The latter part originates from matching of the MQCD gauge coupling $g_{\mathrm M}$,
as $p_{\mathrm G}=T g_{\mathrm M}^6 c_{\mathrm G}$, with  $c_{\mathrm G}$ being 
a non-perturbative coefficient determined in \cite{Kajantie:2002wa,DiRenzo:2006nh},
and the coupling 
$g_{\mathrm M}^2\sim\gE^2(1+\gE^2/\mE+{\cal O}(\gE^4/\mE^2))\sim T\gR^2(1+\gR+{\cal O}(\gR^2))$ 
with coefficients known from e.g.\ \cite{Farakos:1994kx}.

The contributions of order $\gR^7$ to the pressure of hot QCD from the soft scale $gT$
entering through EQCD originate from a number of different sources. 
Within EQCD, the expansion reads (see e.g.\ \cite{Braaten:1995jr})
\begin{align}
p_{\mathrm{M}}(T) &= T \mEr^3 
\lk \frac{\Nc^2\!-\!1}{12\pi} 
+ a_2\,x
+ a_2'\,y
+ a_3\,x^2
+ a_3'\,x\,y
+ a_4\,x^3
+ a_5\,x^4 +\dots\rk\,,
\end{align}
with known 2..4-loop coefficients $a_{2..4}$ \cite{Kajantie:2003ax}, and $a_5$ coming from 
a 5-loop calculation of the EQCD pressure including contributions from
higher-order operators \cite{Chapman:1994vk}
omitted in \eq\nr{eq:eqcd_lagr}, both of which remain unknown to date.
The expansion parameters above relate to the 4d coupling as
\begin{align}
x\equiv\frac{\gEr^2}{\mEr} \sim \gR(1+{\cal O}(\gR^2)) \quad,\quad
y\equiv\frac{\lbothr}{\mEr} \sim \gR^3(1+{\cal O}(\gR^2)) \;,
\end{align}
such that we can already fix the other $g^7$ pieces by multiplying out the 
expansion parameters and our new 3-loop result for $\mE^2$.
Writing
\begin{align}
\mEr^2 &= 
\gR^2\,T^2 \lk \aEfour 
+ \frac{\gR^2}{(4\pi)^2}\,\aEsix
+ \frac{\gR^4}{(4\pi)^4}\,\aEeight +\dots\rk \;,
\end{align}
where $\aEeight$ represents our new 3-loop result of \eq\nr{eq:final},
this coefficient contributes a $g^7$ term to the QCD pressure, via
\begin{align}
\mEr^3&=\gR^3\,T^3\,\aEfour^{3/2}\lk 1 
+\frac{\gR^2}{(4\pi)^2}\,\frac32\,\frac{\aEsix}{\aEfour}
+\frac{\gR^4}{(4\pi)^4}\,\frac38\,\frac{\aEsix^2+4\aEfour\aEeight}{\aEfour^2}
+\dots\rk\;,\\
p_{\mathrm{M}}(T)|_{\mEr^3,\gR^7}&=T^4\frac{\gR^7}{(4\pi)^5}\,\frac{\Nc^2\!-\!1}{8}\,\frac{\aEsix^2+4\aEfour\aEeight}{\aEfour^{1/2}}\nonumber\\
&=8\pi^2 T^4 (\Nc^2\!-\!1) \lk\frac{\Nc}{3}\,\gren\rk^{\frac72} 
\lk 605 L^2 +479 L -180 L_3 + \frac{1487}{4} - 18\zeta(3) \rk\;,
\end{align}
with logarithms $L$ and $L_3$ as in \eq\nr{eq:final} above.

%
\section{Conclusions}
\la{se:conclu}

In this paper, we have determined the Debye screening mass of hot 
Yang-Mills theory to 3-loop (NNLO) accuracy,
by combining various results of a long-term project,
with a number of independent ingredients, 
each of which needed novel state-of-the art techniques
for a successful determination.
While the feasibility of such a precision calculation in the thermal
context was not at all clear from the outset, we have here succeeded
to overcome the last major obstacle -- namely to map a sum of
seven non-trivial master sum-integrals (that remained after 
the IBP reduction algorithm had halted, but which would have been 
needed to an expansion depth for which no technique existed)
to a small number of (three) computable cases. We have achieved
this reduction in complexity of the calculation by a clever basis
transformation, which was made possible by searching our extensive 
database of IBP relations.

To our utmost satisfaction, the final assembly of all building blocks
revealed (a) gauge parameter independence, (b) a finite result after
renormalization, and (c) good convergence properties.
We were therefore able to add another term to the pool of known 
(and heavily used) matching coefficients of EQCD, and to utilize
this fresh term right away, determining one of the {\em physical
next-to-leading order} ($g^7$) contributions to the pressure
of hot QCD. As we have discussed above, this is of course not the 
complete $g^7$ result, but represents a well-defined (and gauge invariant)
contribution to it.

Thus, looking back on the many technical and systematic advances that 
have been made during this project, we conclude that a determination
of the 3-loop effective gauge coupling $\gE^2$, which originates from 
a (by one) higher moment of two-point functions and which
is hence amenable to the same techniques as $\mE^2$,
should be within reach.

Another avenue for further investigations would be a generalization
of our strategy to fermionic contributions, which we have ignored
completely here, setting $\Nf=0$.
While the reduction to basketball-type master integrals is
done \cite{Moeller:2012da},
open problems are finding a suitable basis change,
and evaluating the corresponding fermionic masters -- which, containing
no zero-modes on fermionic lines, could however turn out to be
less involved than the bosonic cases that we have used here.

The result for the Debye mass shows a good convergence in a large temperature range, suggesting that already at the three-loop order the corrections are numerically negligible, but would serve merely in future perturbative calculations (of observables such as the pressure of QCD) to ensure finite renormalized results (cancellation of UV divergences). In the light of the apparent fast convergence of the analytic result, a re-evaluation of the non-perturbative constant of the QCD screening mass can be considered \cite{Laine:1999hh,Burnier:2015nsa} since this latter quantity is in fact used in studies of quark gluon plasma parameters such as the jet quenching parameter \cite{Ghiglieri:2015zma,Panero:2014sua}.

\acknowledgments

We thank K.~Kajantie and M.~Laine for helpful discussions.
Our work has been supported in part by the 
DFG grant GRK~881,
SNF grant 200021-140234, Academy of Finland, grant 27354 
and Magnus Ehrnrooth Foundation (I.G.), 
BMBF project 06BI9002 and DFG grant SCHR~993/2 (J.M.)
as well as DFG grant SCHR~993/1, FONDECYT project 1151281
and UBB project GI-152609/VC (Y.S.).
Our diagrams were drawn with Axodraw~\cite{Vermaseren:1994je}.


\begin{appendix} 

%
\section{Lower-order ingredients}
\la{se:lower}

The one- and two-loop expressions entering \eq\nr{eq:mel}
have been given in Ref.~\cite{Moeller:2012da} in terms of one-loop
sum-integrals $I$. Their $\Nf=0$ pieces read
\begin{align}
\Pi_{\mathrm E1} &= \Nc(d-1)^2\,I_1\;,\\
\Pi_{\mathrm E1}' &= -\Nc\lk\frac{28-5d+d^2}6+(d-3)\xi\rk I_2\;,\\
\Pi_{\mathrm E1}'' &= \Nc\lk\frac{46-7d+d^2}{30}+\frac{d-3}3\,\xi
+\frac{d-6}{12}\,\xi^2\rk I_3\;,\\
\Pi_{\mathrm E2} &= -\Nc^2(1+\xi)(d-3)(d-1)^2\,\Iba\;,\\
\Pi_{\mathrm E2}' &= \Nc^2\lk-\frac{72-42d+13d^2-d^3}{3d(d-7)}
+\frac{d}3\,\xi+\frac{d-6}6\,\xi^2\rk (d-1)^2\,\Ica
+\\&+
\Nc^2\lk\frac{p(d)}{2d(d\!-\!7)(d\!-\!5)(d\!-\!2)}
-\frac{44\!-\!29d\!+\!7d^2\!-\!d^3}{6(d-2)}\,\xi
+\frac{16\!-\!13d\!+\!3d^2}{8(d-2)}\,\xi^2\rk(d\!-\!3)\,\Ibb
\nonumber\;,
\end{align}
with $p(d)=56+315d-231d^2+57d^3-5d^4$.

%
\section{Master sum-integrals}
\la{se:masters}

Let us define a generic notation for
massless 3-loop vacuum sum-integrals
\begin{align}\la{eq:Jdef}
J_{abcdef}^{\alpha\beta\gamma} \equiv 
\mercedes \equiv 
\sumint{PQR} 
\frac{(P_0)^{\alpha} (Q_0)^{\beta} (R_0)^{\gamma}}
{[P^2]^a [Q^2]^b [R^2]^c [(P-Q)^2]^d [(P-R)^2]^e [(Q-R)^2]^f}\;,
\end{align}
where all momenta are understood bosonic.
Let us remark that from the outset, 
only integrals $J^{000}_{abcdef}$ enter the calculation; however, due
to the fact that the IBP relations act in the $d$ spatial dimensions
only and hence explicitly break $(d\!+\!1)$\/-dimensional rotational
invariance introducing the 4-vector $U=(1,\vec 0)$, the numerator
structure of \eq\nr{eq:Jdef} occurs naturally in the reduction step.
The original integral reduction \eq\nr{eq:PiOld} contains 
only basketball-type sum-integrals $J_{ab00ef}^{\alpha\beta\gamma}$
as well as trivial products of one-loop cases
\begin{align}
\la{eq:JIdef}
J_{abc000}^{\alpha\beta\gamma} &=
I_a^\alpha\,I^\beta_b\,I^\gamma_c \;,\\
\la{eq:Idef}
I_s^a &\equiv \sumint{Q} \frac{|\qz|^a}{[Q^2]^s} 
= \frac{2T\,\zeta(2s-a-d)}{(2\pi T)^{2s-a-d}}\,
\frac{\Gamma(s-\frac{d}2)}{(4\pi)^{d/2}\Gamma(s)}
\;,\quad
I_s \equiv \sumint{Q} \frac1{[Q^2]^s} 
= I_s^0\;.
\end{align}
Of these, we need the products
\begin{align}
\la{eq:IIIdef}
\Icaa &= 
\frac{T^2}{(4\pi)^4}
\(\frac{e^\gamma}{4\pi T^2}\)^{3\e}
\frac{2\,\zeta(3)}{144}
\bigg[ 1 +2\e(3-3\gammaE+Z_3+2Z_1) +\order{\e^2} \bigg] \;,\\
\Ibba &= 
\frac{T^2}{(4\pi)^4}
\(\frac{e^\gamma}{4\pi T^2}\)^{3\e}
\frac{1}{12\,\e^2}
\bigg[ 1 +2\e(1-\gammaE+Z_1) 
+2\e^2(2+\tfrac{3\pi^2}8+2Z_1
-\nn&-
\gammaE(\gammaE+2+2Z_1)-4\gamma_1
+\tfrac{\zeta''(-1)}{\zeta(-1)}) +\order{\e^3} \bigg] \;,
\end{align}
containing the numbers $\gammaE$ and $\gamma_1$ arising from the expansion 
of the Riemann Zeta function around its pole at unity, 
$\zeta(1-\e)\approx-1/\e+\gammaE+\gamma_1\e+\dots$,
as well as $Z_n\equiv\zeta'(-n)/\zeta(-n)$.
The specific cases on the left-hand sides of 
\eqs\nr{eq:basistrafo1}-\nr{eq:basistrafo3} are more complicated
integrals, which have however already been evaluated
up to their constant terms in a number of tour de force
computations documented in
Refs.~\cite{Andersen:2008bz,Schroder:2012hm}, 
\cite{Ghisoiu:2012kn} and 
\cite{Ghisoiu:2012yk}, respectively,
from where we collect the results for convenience\footnote{Note that in 
those references, the naming scheme is 
$\{\intV,\intVb,\intMcb\}=\{{\cal M}_{1,0},V_2,{\cal M}_{3,-2}\}.$}:
\begin{alignat}{2}
\intV &\equiv \Jaa &\;=\;& 
\frac{T^2}{(4\pi)^4}
\(\frac{1}{4\pi T^2}\)^{3\e}
\frac{-1}{4\,\e^2}
\lk 1 
+\!\(\frac43\!+\!\gammaE\!+\!2Z_1\)\e
+\!c_1\e^2 
+\!\order{\e^3}\rk \;,\\
\intVb &\equiv \Jab &\;=\;& 
\frac{T^2}{(4\pi)^4} 
\(\frac{1}{4\pi T^2}\)^{3\e}
\frac{1}{96\,\e^2} 
\lk 1 
+\!\(\frac{67}{6}\!+\!\gammaE\!+\!2Z_1\)\e 
+\!c_2\e^2 
+\!\order{\e^3}\rk \;,\\
\intMcb &\equiv \Jac &\;=\;& 
\frac{T^2}{(4\pi)^4}
\(\frac{1}{4\pi T^2}\)^{3\e}
\frac{-5}{36\,\e^2}
\lk 1
+\!\(\frac{71}{30}\!+\!\gammaE\!+\!2Z_1\)\e 
+\!c_3\,\e^2
+\!\order{\e^3} \rk \;.
\end{alignat}
The constant parts are known numerically only, 
and have been determined in the above-mentioned references to be
$c_1\approx+43.8676(1)$, 
$c_2\approx+93.0894417(2)$ and 
$c_3\approx+44.629857(1)$.
In the present computation, however, these constant parts do not 
contribute, since the integrals are multiplied by pre-factors 
$\sim\e$, cf.\ \eq\nr{eq:PiNew}.

%
\section{Coefficients of \eq\nr{eq:PiOld}}
\la{se:coeffs}

The coefficients $\alpha_{1...10}$ of \eq\nr{eq:PiOld}, as
determined in Ref.~\cite{Moeller:2012da}, are rational functions
in $d$ (recall that we use $d=3-2\e$ in this work), 
in some cases also containing the gauge parameter $\xi$,
\begin{align}
\alpha_1 &= -\frac{(d-1) a_1(d)}{24 (d - 6) (d - 5) (d - 4) (d - 3) (d - 2)
(3 d - 17)} \;,\nn
\alpha_2 &= -\frac{3 (d-1) a_2(d)}{4 (d - 6) (d - 5) (d - 3) (d - 2) (3 d -
17)} \;,\nn
\alpha_3 &= -\frac{(d-1)a_3(d)}{4 (d - 6) (d - 5) (d - 2) (3 d - 17)} \;,\nn
\alpha_4 &= -\frac{(d-1)a_4(d)}{12 (d - 6) (d - 5) (d - 3) (d - 2) (3 d -
17)} \;,\nn
\alpha_5 &= \frac{36 (d -11) (d -9) (d -1)}{ (d -6) (d -2) (3d -17)} \;,\nn
\alpha_6 &= \frac{512(d-1) (1954 - 641 d + 48 d^2)}{(d - 5) (d - 3) (d -
2)} \;,\nn
\alpha_7 &= -\frac{49152 (d-1)}{(d-5) (d - 3) (d - 2)} \;,\nn
\alpha_8 &= -\frac{122880 (d -1)}{(d-5) (d - 3) (d - 2)} \;,\nn
\alpha_9 &= -\frac{(d -1) a_9(d)}{4 (d - 6) (d - 5) (d - 3) (d - 2) (3 d -
17)} 
+(d-1)^4\xi+\frac{(d-6)(d-1)^4}{12}\,\xi^2
\;,\nn
\alpha_{10} &= -\frac{(d -1) a_{10}(d)}{4 (d - 7) (d - 6) (d - 5)^2 (d - 3)
(d - 2) (3 d - 17)}
\nn
&+\frac{(d-3) (d-1)^2 (16 - 13 d + 2 d^2)}{2 (d-2)}\,\xi
+\frac{(d-3) (d-1)^2 (16 - 13 d + 3 d^2)}{8 (d-2)}\,\xi^2 \;,
\end{align}
where we have for convenience used the abbreviations
\begin{align}
a_1(d) &= 5982874650 - 9764062527 d + 6860483170 d^2 - 2710270726 d^3 +
 658312418 d^4 \nn
&- 100632587 d^5 + 9447810 d^6 - 497520 d^7 + 11232 d^8, \nn
a_2(d) &= 14947857 - 14330519 d + 5758990 d^2 - 1245506 d^3 + 153345 d^4 -
10215 d^5 + 288 d^6, \nn
a_3(d) &= 8970183 - 7006766 d + 2180196 d^2 - 337402 d^3 + 25941 d^4 - 792
d^5, \nn
a_4(d) &= 1784823003 - 1809632517 d + 757032878 d^2 - 167080938 d^3 +
 20490319 d^4 \nn
&- 1321425 d^5 + 34920 d^6, \nn
a_9(d) &= 17888670 - 22432867 d + 10547330 d^2 - 2313976 d^3 + 186752 d^4 +
 17787 d^5 \nn
&- 5140 d^6 + 416 d^7 - 12 d^8, \nn
a_{10}(d) &= - 835002999 + 1120616178 d - 653300169 d^2 + 218578438 d^3 -
 47171745 d^4 \nn
&+ 7193512 d^5 - 870355 d^6 + 92054 d^7 - 8000 d^8 +
 458 d^9 - 12 d^{10}.
\end{align}

%
\section{IBP relations for basis transformation}
\la{se:ibp}

The idea of performing a basis transformation on $\Pi_{\mathrm E3}$ 
translates into going some steps back into its IBP reduction 
\cite{Moeller:2012da}. 
The goal is to search for relations that change the coefficients of the 
master sum-integrals (cf.\ the last paragraph of \app{C} in \cite{Moeller:2012da})
in such a way as to eliminate all factors of $(d-3)$ in the denominator.
While it is not at all clear from the outset that this can always be
achieved, it happens indeed if we use the following three automatically 
generated IBP relations, expressed in terms of the 
basis of basketball-type 3-loop master integrals 
defined in \eq\nr{eq:Jmasters}:
\begin{align}
\la{eq:basistrafo1}
\intV &= 
 \frac{2(47-24d+3d^2)}{3(d-3)^2(d-4)}\,J_1
 +\frac{16}{3(d-3)^2}\,J_4 \;,\\
\la{eq:basistrafo2}
\intVb &= 
\frac{(d-9)(d-7)(d-2)}{2(d-6)(d-5)(d-4)(d-3)^2}\,\Icaa 
+\frac{519-312d+61d^2-4d^3}{2(d-6)(d-5)^2(d-4)(d-3)}\,\Ibba
-\nn&-
\frac{(3d-10)(10791-9060d+2806d^2-380d^3+19d^4)}
  {12(d-6)(d-5)(d-4)^2(d-3)^2}\,J_1
+\nn&+
\frac{3(d-7)}{2(d-6)(d-4)(d-3)}\,J_2
 -\frac{(d-9)(d-7)}{2(d-6)(d-5)(d-4)(d-3)}\,J_3
+\nn&+
\frac{31401-16707d+2951d^2-173d^3}{6(d-6)(d-5)(d-4)(d-3)^2}\,J_4
 +\frac{512}{(d-5)(d-4)(d-3)^2}\,J_6
\;,\\
\la{eq:basistrafo3}
\intMcb &=
\frac{b_1(d)\,J_1+18(d-4)b_2(d)\,J_2+6(d-4)(d-3)b_3(d)\,J_3+2(d-4)b_4(d)\,J_4}
{12(d-7)(d-6)(d-5)(d-4)(d-3)^2(d-2)(d-1)(3d-17)}
-\nn&-
\frac{72(d-11)(d-9)\,J_5}{(d-7)(d-6)(d-3)(d-2)(d-1)(3d-17)}
+\nn&+
1024\,\frac{(-1970+665d-56d^2)\,J_6 +96\,J_7 +240\,J_8}
 {(d-7)(d-5)(d-3)^2(d-2)(d-1)}
+\nn&+
\frac{(d-7)(d-5)b_9(d)\,\Icaa +b_{10}(d)\,\Ibba}
  {2(d-7)^2(d-6)(d-5)^2(d-3)^2(d-2)(d-1)(3d-17)}\;,
\end{align}
where \eq\nr{eq:basistrafo3} contains the polynomials
\begin{align}
b_1(d) &= 6039084810-9921665183d+7037865926d^2-2817068438d^3
       +696438686d^4\nn&-108972587d^5+10546622d^6-577632d^7+13716d^8\;,\nn
b_2(d) &= 14890737-14196135d+5642398d^2-1196706d^3+142785d^4-9079d^5
       +240d^6\;,\nn
b_3(d) &= 8935911-6940606d+2138852d^2-326810d^3+24757d^4-744d^5\;,\nn
b_4(d) &= 1801692987-1846811125d+786404686d^2-178072266d^3+22616879d^4\nn&
       -1527809d^5+42888d^6\;,\nn
b_9(d) &= 18419886-24034307d+12505438d^2-3593336d^3+677116d^4-95461d^5\nn&
        +10336d^6-736d^7+24d^8\;,\nn
b_{10}(d) &= -834536043+1116882072d-645264933d^2+210429806d^3
          -42367633d^4\nn&+5387676d^5-421763d^6+18526d^7-348d^8\;.
\end{align}


\end{appendix}



\end{document}